\documentclass[12pt]{article}

\usepackage{amsmath, amssymb, slashed}
\usepackage[dvips]{graphicx}
\usepackage{epsf}

\newcommand{\psl}{\mathbf{p} \hspace{-0.5 em}/}
\newcommand{\psll}{\mathbf{P} \hspace{-0.5 em}/}
\newcommand{\ptsl}{p \hspace{-0.5 em}/_T}

\begin{document}

\begin{titlepage}
\begin{center}

\hfill UT-11-38 \\
\hfill TU-894 \\
\hfill IPMU-11-0188 \\

\vspace{1.0cm}
{\large\bf Exploring Supersymmetric Model \\
with Very Light Gravitino at the LHC}
\vspace{2.0cm}

{\bf Masaki Asano}$^{(a, b)}$,
{\bf Takumi Ito}$^{(b, c)}$,
{\bf Shigeki Matsumoto}$^{(d)}$
\\
and
{\bf Takeo Moroi}$^{(b, d)}$

\vspace{1.0cm}
{\it
$^{(a)}${\it II. Institute for Theoretical Physics, 
University of Hamburg, \\
Luruper Chausse 149, DE-22761 Hamburg, Germany} \\
$^{(b)}${\it Department of Physics, University of Tokyo, 
Tokyo 113-0033, Japan} \\
$^{(c)}${\it Department of Physics, Tohoku University, 
Sendai 980-8578, Japan} \\
$^{(d)}${\it IPMU, TODIAS, University of Tokyo, 
Kashiwa, 277-8583, Japan}
}
\vspace{2.0cm}

\abstract{ The low-scale gauge mediation scenario of supersymmetry
  breaking predicts very light gravitino, which makes the next
  lightest supersymmetric particle (NLSP) quasi stable.  We study the
  LHC phenomenology of the case that the NLSP is the stau. When the
  mass of the gravitino is of the order of 10eV, the decay length of
  stau is about 0.1--1~mm, so that it decays before reaching the first
  layer of the inner silicon detector. We show, however, that, with
  utilizing the impact parameter of $\tau$-jets from the stau decay,
  it is possible to determine the mass spectrum of sparticles
  precisely. It is also possible to estimate the lifetime of the stau
  by observing distribution of the impact parameter.  }

\end{center}
\end{titlepage}
\setcounter{footnote}{0}

\section{Introduction}
\label{sec: intro}

Large Hadron Collider experiment (LHC) is now operating and
reports many important results on new physics beyond the standard
model (SM). 
Although positive signals have not been reported so far,
those are expected to be found in near future,
because the hierarchy problem of the SM strongly suggests the existence of new
physics at the TeV scale or below. On the other hand, many new physics
models have been theoretically proposed. 
Among those, the supersymmetric model is very attractive because it guarantees
the stability of the Higgs mass to its radiative corrections and gives
a clue to solve the hierarchy problem. In addition, the supersymmetry
(SUSY) plays a crucial role to realize the grand unification of known
gauge interactions of the SM at a certain high energy scale.

Details of supersymmetric model, such as the mass spectrum of
sparticles, depend highly on how SUSY is broken.  So far, a variety of
SUSY breaking mechanisms has been proposed~\cite{BookDrees}.  Among
those, the gauge mediation scenario~\cite{Dine:1993yw} attracts an
attention, because it gives a solution to dangerous SUSY flavor
problems. In this scenario, the breaking occurs at lower energy scale
than those of other SUSY breaking scenarios, so that the superpartner
of graviton, the gravitino, is likely to be the lightest
supersymmetric particle (LSP). The gravitino mass is predicted to be
in the range between ${\cal O}(10)$eV and ${\cal O}(1)$GeV.  In this
article, we focus on the low-scale gauge mediation model providing a
gravitino with ${\cal O}(10)$eV mass.  Such a scenario is well
motivated because it is completely free from severe cosmological
constraints~\cite{Feng:2010ij} such as Big-Bang
Nucleosynthesis~\cite{Kawasaki:2008qe} and large scale structure
formation of our universe~\cite{Viel:2005qj}.

Collider signals of the low-scale gauge mediation scenario depend on
what the next lightest superparticle (NLSP) is, which decays only into
gravitino and its superpartner. Though there are many candidates for
NLSP, we focus on the stau NLSP in this article, which is predicted in
wide parameter region of the scenario. When the gravitino mass is of
${\cal O}(10)$eV, the stau NLSP decays into a $\tau$-lepton and a
gravitino with the lifetime of $10^{-15}$--$10^{-11}$sec. The decay
length (the lifetime times the speed of light) of the stau NLSP is
therefore much shorter than the typical size of collider detectors,
and the traditional supersymmetric signal, namely, multi-jets
associated with missing energy and $\tau$-leptons, is expected at the
LHC experiment. Such a signal is, however, generally predicted in
various SUSY breaking scenarios.

We show in this article that, even if the NLSP decays before reaching
inner trackers of collider detectors, we can use the impact-parameter
information about the decay products of the NLSP to study various
properties of superparticles.  In particular, the impact parameter is
available for charged tracks caused by decay products of $\tau$-lepton
at the stau NLSP decay.  If the decay product of stau is found to have
large impact parameter, it strongly suggests that the underlying SUSY
breaking scenario is low-energy gauge mediation. Furthermore, the
impact parameter is also utilized to precisely measure the spectrum of
sparticles such as squark, neutralino, and stau masses. This is
because two tau leptons produced by the cascade decay of a squark can
be distinguished with each other by using the impact parameter.  In
addition, we may be able to determine the lifetime of the NLSP (i.e.,
stau in the present study) using the impact parameter distribution.
When mass and lifetime of the stau NLSP are measured, it is possible
to determine the gravitino mass assuming that the stau decays into
gravitino and tau. The scale of SUSY breaking in the low-energy gauge
mediation scenario is, therefore, obtained.  It has been already shown
that such studies can be easily performed once the $e^+e^-$ linear
collider becomes available \cite{Matsumoto:2011fk}.  Here, we consider
the case of the LHC.  We will see that the measurement of the mass
spectrum as well as the determination of the lifetime of the NLSP can
be performed at the LHC with the help of impact parameter information.

This article is organized as follows. In the next section, we consider
some properties of the stau NLSP in the low-energy gauge mediation
scenario and discuss how the impact parameter from the NLSP decay is
utilized in determinations of sparticle masses and NLSP lifetime. Our
simulation framework is summarized in section \ref{sec: framework}, in
which a representative point and several strategies to reduce
backgrounds are shown. In section \ref{sec: results}, simulation
results for the measurements of sparticle masses and lifetime of the
stau NLSP are discussed. Section \ref{sec: summary} is devoted to
summary of our studies.

\section{Utilizing impact parameter}
\label{sec: impact parameter}

In this section, we discuss how the impact parameter is utilized in
order to determine the mass spectrum of sparticles and the lifetime of
stau NLSP. We first briefly review some properties of the stau NLSP
and define the impact parameter.  Then, we discuss basic strategies
for the measurement of the mass spectrum and the lifetime of NLSP with
the use of the impact parameter.

\subsection{Stau NLSP and impact parameter}

Since the LHC is a hadron collider, colored sparticles such as squarks
and gluino are expected to be produced at first, which decay into the
stau NLSP, the super partner of $\tau$-lepton, through several cascade
channels. The stau NLSP then decays into a $\tau$-lepton and a
gravitino with the following lifetime,
\begin{eqnarray}
\tau_{\tilde \tau}
=
48 \pi M_{\rm pl}^2 \left(\frac{m_{3/2}^2}{m_{\tilde \tau}^5}\right)
\simeq
5.9 \times 10^{-12}~[{\rm sec}]
\times
\left( \frac{m_{3/2}}{10{\rm eV}} \right)^2
\left( \frac{100{\rm GeV}}{m_{\tilde \tau}} \right)^5,
\label{eq: lifetime}
\end{eqnarray}
where $M_{\rm pl} \simeq 2.4 \times 10^{18}$GeV, $m_{\tilde \tau}$,
and $m_{3/2}$ are reduced Planck mass, stau mass, and gravitino mass,
respectively. It turns out from above formula that the decay length
(the lifetime $\times$ the speed of light) of the stau NLSP is
estimated to be $\sim {\cal O}(100)\mu$m when gravitino and stau
masses are $\sim 10$eV and $\sim 100$GeV, respectively. On the other
hand, the decay length of $\tau$-lepton which is one of main
backgrounds against the stau signal, is 87$\mu$m, so that the decay of
stau NLSP into very light gravitino can be, in principal, detected if
we can reduce SM backgrounds efficiently.

Since the decay length of the stau NLSP is, at most, ${\cal O}(1)$mm
in the parameter region of our interest, the stau NLSP decays before
reaching the first pixel detector, which is located at 5cm (4cm) away from
the beam line in the ATLAS detector~\cite{Aad:2009wy} (CMS detector~\cite{Bayatian:2006zz}).
The lifetime of the stau NLSP is, as a result, difficult to be
measured using methods usually applied to detect long-lived particles,
such as methods by observing charged tracks \cite{Ishiwata:2008tp,
  Kaneko:2008re, Asai:2008sk, Asai:2011wy}.  On the other hand, the
lifetime of the stau NLSP may still be possible to be determined using
the distribution of the impact parameter, which is obtained by
$\tau$-jets from the stau NLSP decay. The impact parameter is defined
as the shortest distance to the track from the interaction point. The
positional resolution of the ATLAS detector along the longitudinal
direction is $\Delta_L \sim 100 \mu$m, which is not good compared to
that along the transverse direction, $\Delta_T \sim 10
\mu$m.\footnote{
  Details of those performances are found in the section "Tracking" in
  Ref.~\cite{Aad:2009wy}.
}
Thus, we use the transverse impact parameter which is defined by
\begin{eqnarray}
  d_I \equiv
  \left|
    {\bf x}_T^I - 
    \frac{{\bf x}_T^I \cdot {\bf P}_T^I}{|{\bf P}_T^I|^2}{\bf P}_T^I
  \right|,
  \label{eq: impact parameter}
\end{eqnarray}
where ${\bf x}_T^I$ and ${\bf P}_T^I$ are transverse decay point of
the $I$-th $\tau$-lepton and transverse momentum of the tau-jets from
the $I$-th $\tau$-lepton decay, respectively. Note that the summation
over the index $I$ should not be taken here. The above formula is used
in our simulation studies, which will be presented in following
sections. We expect that, at the LHC experiment, the distribution of
the impact parameter $d_I$ is obtained by measuring the shortest
distance (projected onto the transverse-plane) to the $\tau$-jet track
from the interaction point.

\subsection{Impact parameter for mass measurements}
\label{sec: IP for masses}

We next consider how the impact parameter is utilized in mass
measurements of sparticles. At the LHC, colored sparticles such as
gluino and squarks are expected to be produced copiously, and
non-colored sparticles are then produced through cascade decays of the
colored ones. The chain of the cascade decay is, for example, composed
of following processes; First colored sparticle decays into a
neutralino/chargino by emitting a quark which is observed as a
jet. Next a neutralino/chargino decays into a slepton by emitting a
lepton. Finally, a slepton decays into a LSP by again emitting a
lepton. It is needless to say that the LSP passes through the detector
without giving any signatures, which is, instead, observed as a
missing energy. This chain (called "golden mode") is frequently used
to measure the mass spectrum of sparticles in various supersymmetric
scenarios by using several kinematical
endpoints~\cite{Hinchliffe:1996iu, Hinchliffe:1998ys}.

\begin{figure}[t]
\begin{center}
\includegraphics[origin=b, angle=0, width=8cm]{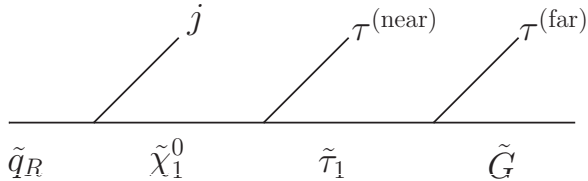}
\caption{\small Typical decay chain in the low-scale gauge mediation scenario.} 
\label{fig: cascade chain}
\end{center}
\end{figure}

In the case of the low-scale gauge mediation scenario, we have a
similar decay chain.  One of the examples is shown in Fig.~\ref{fig:
  cascade chain}, where $\tilde{q}_R$, $\tilde{\chi}^0_1$,
$\tilde{\tau}_1$, and $\tilde{G}$ are right-handed squark, lightest
neutralino, lightest stau, and gravitino, respectively.  The character
$j$ denotes a jet which originates in a quark from the $\tilde{q}_R$
decay.\footnote
{In the gluino production event, the gluino decays into a squark by
  also emitting a jet. Because the mass difference between gluino and
  squark is much smaller than that between squark and neutralino in
  the parameter region of our interest, we can discriminate between a
  (soft) jet from the gluino decay and a (hard) jet from the squark
  decay.}
Following the terminology used in studies of the golden mode, we call
$\tau$ from the $\tilde{\chi}^0_1$-decay $\tau^{\rm (near)}$ and that
from the $\tilde{\tau}_1$-decay $\tau^{\rm (far)}$.  Importantly, the
impact parameter of $\tau^{\rm (far)}$ is expected to be larger than
that of $\tau^{\rm (near)}$, which is of great help for the event
reconstruction.

In our analysis, we focus on the signal from the decay of a squark
shown in Fig.~\ref{fig: cascade chain}. Kinematics of its decay chain
is, as a result, governed by following four sparticle masses; the
masses of squark ($m_{\tilde q}$), lightest neutralino
($m_{\tilde{\chi}_1^0}$), lightest stau ($m_{\tilde{\tau}_1}$), and
gravitino ($m_{3/2}$). Since the gravitino mass is of the order of
10eV, only the upper bound on the mass is expected to be obtained. On
the other hand, the existence of large impact parameters in signal
events strongly suggests the (low-scale) gauge mediation scenario. We
therefore perform our analysis with simply postulating that the
gravitino mass is much smaller than those of other sparticles, namely,
with treating the gravitino as a massless particle. Three independent
kinematical endpoints are then enough to determine the mass spectrum
of sparticles. With the help of the impact parameter, many kinematical
variables are now available. Among those, we use the invariant mass
between two $\tau$-leptons ($M_{\tau^{\rm (near)} \tau^{\rm (far)}}$),
that between jet and near $\tau$-lepton ($M_{j \tau^{\rm (near)}}$),
and the $M_{T2}$ variable from leading two jets ($M_{T2,jj}$).

The upper limit on the invariant mass $M_{\tau^{\rm (near)} \tau^{\rm
    (far)}}$ is given by
\begin{eqnarray}
  M_{\tau^{\rm (near)} \tau^{\rm (far)}}^{\rm max}
  =
  m_{\tilde{\chi}_1^0}
  \sqrt{1 - m^2_{\tilde{\tau}_1}/m^2_{\tilde{\chi}_1^0}},
  \label{eq: mass_tautau}
\end{eqnarray}
where the mass of $\tau$-lepton is set to be zero in above formula. In
addition, the upper limit on the distribution of the invariant mass
$M_{j \tau^{\rm (near)}}$ is given by the following formula,
\begin{eqnarray}
M_{j \tau^{\rm (near)}} ^{\rm max}
=
m_{\tilde{q}}
\sqrt{
\left(1 - m^2_{\tilde{\chi}_1^0}/m^2_{\tilde{q}}\right)
\left(1 - m^2_{\tilde{\tau}_1}/m^2_{\tilde{\chi}_1^0}\right)}.
\label{eq: mass_jtau}
\end{eqnarray}
The last kinematical variable used in our analysis is the $M_{T2}$
variable~\cite{Lester:1999tx} constructed from highest two jets; denoting the
momenta of highest two jets as ${\bf p}$ and ${\bf p}'$, we define
\begin{eqnarray}
  M_{T2,jj}(m_{\rm miss})
  =
  \min_{{\bf k}_T + {\bf k}_T^{\prime} = {\psll}^{\rm eff}_T}
  \left[
    \max 
    \left\{ M_T({\bf p}_T, {\bf k}_T), M_T({\bf p}'_T, {\bf k}'_T) \right\}
  \right],
\end{eqnarray}
where $M_T$ is the transverse mass and $m_{\rm miss}$ is the ``test
mass.''  Here, because we construct the $M_{T2}$ variable only from
highest two jets, ${\psll}^{\rm eff}_T$ should be understood as the
vector sum of transverse momenta of all the activities other than
highest two jets and missing momentum $\psl_T$: ${\psll}^{\rm eff}_T =
\psl_T + \sum_i {\bf p}_{\tau-jet~i} + \sum_i {\bf
  p}^{\prime}_{\tau-jet~i}$.  The upper limit of this variable is then
given by
\begin{eqnarray}
  M_{T2,jj}^{\rm max}(m_{\rm miss})
  =
  \frac{m^2_{\tilde{q}} - m^2_{\tilde{\chi}_1^0}}{2m_{\tilde{q}}}
  +
  \sqrt{
    \left(
      \frac{m^2_{\tilde{q}} - m^2_{\tilde{\chi}_1^0}}{2m_{\tilde{q}}}
    \right)^2 + m_{\rm miss}^2
  }.
  \label{eq: MT2_jj}
\end{eqnarray}

Using three kinematical endpoints given in eqs.(\ref{eq:
  mass_tautau}), (\ref{eq: mass_jtau}) and (\ref{eq: MT2_jj}), we fit
the sparticle masses $m_{\tilde q}$, $m_{\tilde{\chi}_1^0}$, and
$m_{\tilde{\tau}_1}$. We will see that, though the cascade chain
always involves $\tau$s as lepton emissions, the spectrum can be
determined accurately because of information about the impact
parameter.

\subsection{Impact parameter for stau lifetime measurement}
\label{sec: IP for lifetime}

Information about the lifetime of the stau NLSP is imprinted in the
distribution of the impact parameter. The impact parameter, however,
depends not only on the lifetime of the stau NLSP but also on its mass
and velocity.  With the use of the strategy discussed in previous
subsection, the stau mass is measured precisely. On the other hand,
since the gravitino produced from the stau decay cannot be detected,
the velocity cannot be determined on event-by-event basis, which makes
it difficult to determine the lifetime using the impact parameter
distribution.

Once the mass spectrum of the superparticles are known, however, we
expect to acquire information about the velocity distribution of the
$\tilde{\tau}_1$ in supersymmetric events.  In the present case (where
the mass spectrum of a simple gauge mediation model is assumed), we
may understand that the underlying scenario is indeed the low-scale
gauge mediation from the experimentally measured mass spectrum as well
as the confirmation of the existence of long-lived stau.  Even if we
cannot specify the complete structure of the underlying model, we may
still be able to measure the masses of superparticles which are most
important for the determination of the velocity distribution of
$\tilde{\tau}_1$ (i.e., the masses of $\tilde{q}$, $\tilde{\chi}^0_1$,
and $\tilde{\tau}_1$), as we have discussed in the previous section.
Then, once those information becomes available, one will be able to
obtain the velocity distribution with, for example, Monte Carlo
simulation.

In our analysis, we assume that the velocity distribution of
$\tilde{\tau}_1$ can be understood once the superparticles are
discovered.  The detailed study of the methods of determining the
velocity distribution is beyond the scope of this paper, so we simply
assume that the averaged velocity of the produced $\tilde{\tau}_1$ can
be obtained with some accuracy and determine the lifetime using the
$\tilde{\tau}_1$. Although a better determination of the lifetime of
$\tilde{\tau}_1$ may be possible if we can somehow obtain and use the
information about the velocity distribution of $\tilde{\tau}_1$, we
can still have a relatively good determination of the lifetime using
the averaged velocity as we will describe.  The procedure to measure
the lifetime of the stau NLSP is therefore the following.
\begin{enumerate}

\item[(i)] We first assume that the averaged velocity of the stau
  NLSP, denoted as $\bar{\beta}_{\tilde{\tau}_1}$, is somehow
  understood.  Then, we generate $\tilde{\tau}_1$ with the fixed
  velocity $\bar{\beta}_{\tilde{\tau}_1}$ and make signal templates of
  the distribution of the transverse impact parameter $d_I$.  The
  template is prepared for wide range of stau lifetime.  The
  production angle of the stau NLSP is assumed to be isotropic in
  generating the events for the template.

\item[(ii)] Distribution of the impact parameter expected at the LHC
  experiment is obtained by using Monte Carlo simulation.

\item[(iii)] Comparing the templates obtained in the (i) with the
  actual distribution obtained in (ii), we study how well we can
  constrain the lifetime of $\tilde{\tau}_1$ by $\chi^2$--analysis.
  By varying the value of $\bar{\beta}_{\tilde{\tau}_1}$ used in
  making the templates, we also discuss the uncertainty related to the
  determination of the velocity distribution.

\end{enumerate}

Here, we have a few comments on the above method. First comment is on
the effect of gluino production.  The mass difference between gluino
and squark is much smaller than that between squark and neutralino in
the parameter region of our interest.  In addition, it will be
possible to select signal events with the desirable squark decay chain
($\tilde{q} \to \tilde{\chi}^0_1 \to \tilde{\tau}_1 \to \tilde{G}$) by
applying appropriate kinematical cuts.  Thus, the averaged value of
the boost factor is expected to depend weakly on the gluino mass. We
have checked this statement quantitatively by simulating signal events
with several choices of the gluino mass. Second comment is on how the
transverse impact parameter $d_I$ depends on the production angle of
the stau NLSP. One might worry if we may compare the actual impact
parameter distribution with theoretical templates obtained by
postulating isotropic distribution of the production angle. We have
checked that this potential problem can be solved by only using
$\tau$-jets with small pseudo-rapidity.

\section{Simulation framework}
\label{sec: framework}

Before showing our results, we summarize the framework of our
simulation study. We first mention a representative point and
simulation tools used in the study. Next we discuss the strategy to
suppress combinatorial backgrounds of signal events caused by the
existence of two decay chains. We finally consider the SM backgrounds
and discuss kinematical cuts used to reduce those backgrounds.

\subsection{Representative point \& simulation tools}

The representative point used in our simulation study has been chosen
by adopting the minimal model of the gauge mediation symmetry breaking
\cite{Dine:1993yw} with the following underlying parameters; the SUSY
breaking scale ($\varLambda = 30$TeV), the messenger mass scale
($M_{\rm mess} = 300$TeV), the number of SU(5) messenger fields
($N_{\bf 5} = 5$), and the ratio of vacuum expectation values of two
Higgs fields ($\tan\beta = 15$).
The ISAJET package~\cite{ISAJET} is used in order to calculate the
spectrum and branching fractions of sparticles.
Resultant masses and branching
fractions of sparticles relevant to the study are summarized in
Table~\ref{tab: point} for the case of the gravitino mass of 9.7eV which corresponds to the decay length of the stau of 500$\mu$m.
It can be seen that the model is consistent with current LHC data~\cite{Kats:2011qh}.

In our analysis, we consider the LHC experiment with the center of
mass energy of $\sqrt{s}=14{\rm TeV}$.  Then, the signal cross
section, which is the sum of the cross sections for gluino and squark
productions, is estimated to be
\begin{eqnarray}
\sigma_{\tilde{g} \tilde{g}} = 0.129{\rm pb},
\quad
\sigma_{\tilde{q} \tilde{g}} = 0.922{\rm pb},
\quad
\sigma_{\tilde{q} \tilde{q}} = 0.879{\rm pb}.
\end{eqnarray}
We focus on the decay chain involving a right-handed squark, as shown
in Fig.~\ref{fig: cascade chain}. As a result, a typical signal event
consists of two energetic jets, four $\tau$-leptons ($\tau$-jets or
leptons), and a missing energy in the transverse direction.


\begin{table*}[t]
\begin{center}
\begin{tabular}{c|clll}
& Mass (GeV) & \multicolumn{2}{c} {Branching fractions} \\
\hline
$\tilde{g}$ & 1096.6 & Br($\tilde{g} \to \tilde{q} q$) = 0.89. \\
$\tilde{u}_L$ & 951.1 & Br($\tilde{u}_L \to \tilde{\chi}^\pm_2 d$) = 0.34,
                      & Br($\tilde{u}_L \to \tilde{\chi}^\pm_1 d$) = 0.32, & \\
              &       & Br($\tilde{u}_L \to \tilde{\chi}^0_4 u$) = 0.18,
                      & Br($\tilde{u}_L \to \tilde{\chi}^0_2 u$) = 0.15. \\
$\tilde{u}_R$ & 922.0 & Br($\tilde{u}_R \to \tilde{\chi}^0_1 u$) = 0.96. \\
$\tilde{\chi}^0_1$ & 197.3
& Br($\tilde{\chi}^0_1 \to \tilde{\tau}^\pm_1 \tau^\mp$) = 0.35,
& Br($\tilde{\chi}^0_1 \to \tilde{e}^\pm_R e^\mp$) = 0.32, \\
&
& Br($\tilde{\chi}^0_1 \to \tilde{\mu}^\pm_R \mu^\mp$) = 0.32. \\
$\tilde{e}_R$ & 130.0
& \multicolumn{2}{l}
{Br($\tilde{e}_R \to \tilde{\tau}^\pm_1 \tau^\mp e$) $\simeq$ 1.00.} \\
$\tilde{\tau}_1$ & 126.2 & Br($\tilde{\tau}_1 \to \tau \tilde{G}$) = 1.00. \\
\hline
\end{tabular}
\caption{\small Masses and branching fractions of sparticle in our representative point.}
\label{tab: point}
\end{center}
\end{table*}

For parton-level event generation and hadronization, we employ the HERWIG code~\cite{HERWIG,
  HERWIGSUSY}. Generated events are passed through the PGS
code~\cite{PGS4} for simulating detector effects. Fake $\tau$-jets
from QCD processes and heavy meson decays are involved in the
study. For tau-jets, we smear the transverse vertex position of the
parton using Gaussian distribution with the error $\Delta d_I =
10\mu$m.

\subsection{Charge subtraction method}

Background reduction is the most important task in our analysis to
determine the mass spectrum of sparticles, because all kinematical
endpoints do not have sharp edge structures due to the energy leakage
by $\nu_{\tau}$ emissions from $\tau$-decays.  Expected backgrounds in
those measurements are as follows: (i) A number of fake $\tau$-jets
are expected at the hadron collider. (ii) A signal event results in
multiple $\tau$-leptons, and hence there are combinatorial backgrounds
in the analysis involving $\tau$-jets.

In order to reduce these backgrounds, we adopt the method of the
charge subtraction. We expect four $\tau$-leptons (two $\tau^+$ and
two $\tau^-$) in one event. We therefore have three ways to pair the
$\tau$ leptons, $(\tau^\pm, \tau^\mp)_1$, $(\tau^\pm, \tau^\mp)_2$,
and $(\tau^\pm, \tau^\pm)$. In each event (involving four $\tau$s), we
take the data using the method $(\tau^\pm, \tau^\mp)_1 + (\tau^\pm,
\tau^\mp)_2 - (\tau^\pm, \tau^\pm)$, then the wrong opposite-sign
pairing is expected be canceled by the subtraction of the same-sign
pairing. This method works very well when $\tau$-leptons are produced
in the process $\tilde{\chi}^0 \to \tilde{\tau} \to \tilde{G}$; this
is due to the fact that $\tilde{\chi}^0$ decays into
$\tilde{\tau}^+\tau^-$ and $\tilde{\tau}^-\tau^+$ with equal
probability.

In addition, this method can be applied to the determination of the
$M_{j \tau^{\rm (near)}}$-endpoint. We simply collect $(j, \tau,
\tau)$ events using the charge subtraction. In each paring, $\tau^{\rm
  (near)}$ is identified as the $\tau$ lepton which has a shorter
impact parameter. It should be also noted that the charge subtraction
method can reduce backgrounds from fake $\tau$-jets from QCD
processes, because the QCD fake events are charge-blind in a good
approximation at high energy processes.

\subsection{Kinematical cuts to reduce $t\bar{t}$ backgrounds}
\label{sec: selection cuts}

The most serious SM background for our study is the $t\bar{t}$
production. Thus, we concentrate on this background.  In order to
reduce this background, we impose following kinematical cuts:
\begin{itemize}

\item Large missing transverse energy, $\ptsl >$ 150GeV.

\item At least, four leptons, $e$, $\mu$ with $p_T >$ 20GeV or
  $\tau$-jet with $p_T >$ 25GeV.

\item Two hard jets, $j_1$($j_2$) with $p_T >$ 200(100)GeV (and no
  b-jets).

\end{itemize}
Here, we use the label, $i = 1$ or 2, for a jet ($j_i$) in decreasing
order of $p_T$. We also take account of both electrons and muons,
because $\tau$-lepton often decays leptonically and selectron and
smuon decay into gravitino by emitting electron and muon directly. The
requirement for jets to have $p_T >$ 200(100)GeV is very important to
reduce the $t\bar{t}$ background, because jets with $p_T > m_W$ ($m_W$
is the mass of weak gauge boson) from top quark decays are rather
rare. In Table~\ref{table: cut flaw}, we summarize the cut flow in our
simulation study with assuming that the integrated luminosity is
100fb$^{-1}$, where we take the gravitino mass of 9.7eV again.

It is possible to apply tighter kinematical cuts for further
reductions of the backgrounds. For instance, the requirement $\ptsl >$
200GeV, $p_T(j_1) >$ 250GeV and $p_T(j_2) >$ 150GeV in addition to the
basic kinematical cuts shown in Table~\ref{table: cut flaw} will
reduce 80\% of the $t\bar{t}$ background (we have, as a result,
$\sim$400 $t\bar{t}$ events), while this also reduces 30\% of SUSY
signals (we have, as a result, $\sim$28000 SUSY events).

\begin{table}
\begin{center}
\begin{tabular}{l|rr}
Selection cut & SUSY & $t\bar{t}$ \\
\hline
(0) Generated events & 273,600 & 49,610,000 \\
(1) $\ptsl >$ 150GeV & 190,504 & 1,644,160 \\
(2) \# of leptons $\geq$ 4 & 70,641 & 55,537 \\
(3) \# of b-jets = 0 & 56,228 & 41,673 \\
(4) $j_1$ with $p_T >$ 200GeV & 49,819 & 5,915 \\
(5) $j_2$ with $p_T >$ 100GeV & 41,007 & 1,984 \\
\hline
\end{tabular}
\caption{\small Numbers of signal (SUSY) and background ($t\bar{t}$)
  events after applying kinematical cuts with ${\cal L} =$
  100fb$^{-1}$. All SUSY processes are included in our event
  generation.}
\label{table: cut flaw}
\end{center}
\end{table}

\section{Simulation results}
\label{sec: results}

We are now in position to present several results of our simulation
study, which are obtained based on arguments in previous sections. We
first show the results for measurements of sparticle masses, and
discuss how accurately these masses can be determined at the LHC. We
next show that the lifetime of the stau may be determined by using the
distribution of the transverse impact parameter of $\tau$-jets from
the stau NLSP decay. We estimate how accurately the lifetime can be
determined.

\subsection{Sparticle masses} \label{sec: results mass}

The strategy to determine the mass spectrum is the use of kinematical
endpoints of several variables.
We study how the endpoints behaves using generated events which pass through the basic cuts discussed in
section \ref{sec: selection cuts}.
For the simulation study of sparticle mass measurement, the decay length of the stau NLSP is set to be 500$\mu$m
(corresponding to the gravitino mass of 9.7eV).

\subsubsection{Endpoint on $M_{\tau\tau}$}

\begin{figure}[p]
\begin{center}
\includegraphics[origin=b, angle=0, width=8.4cm]{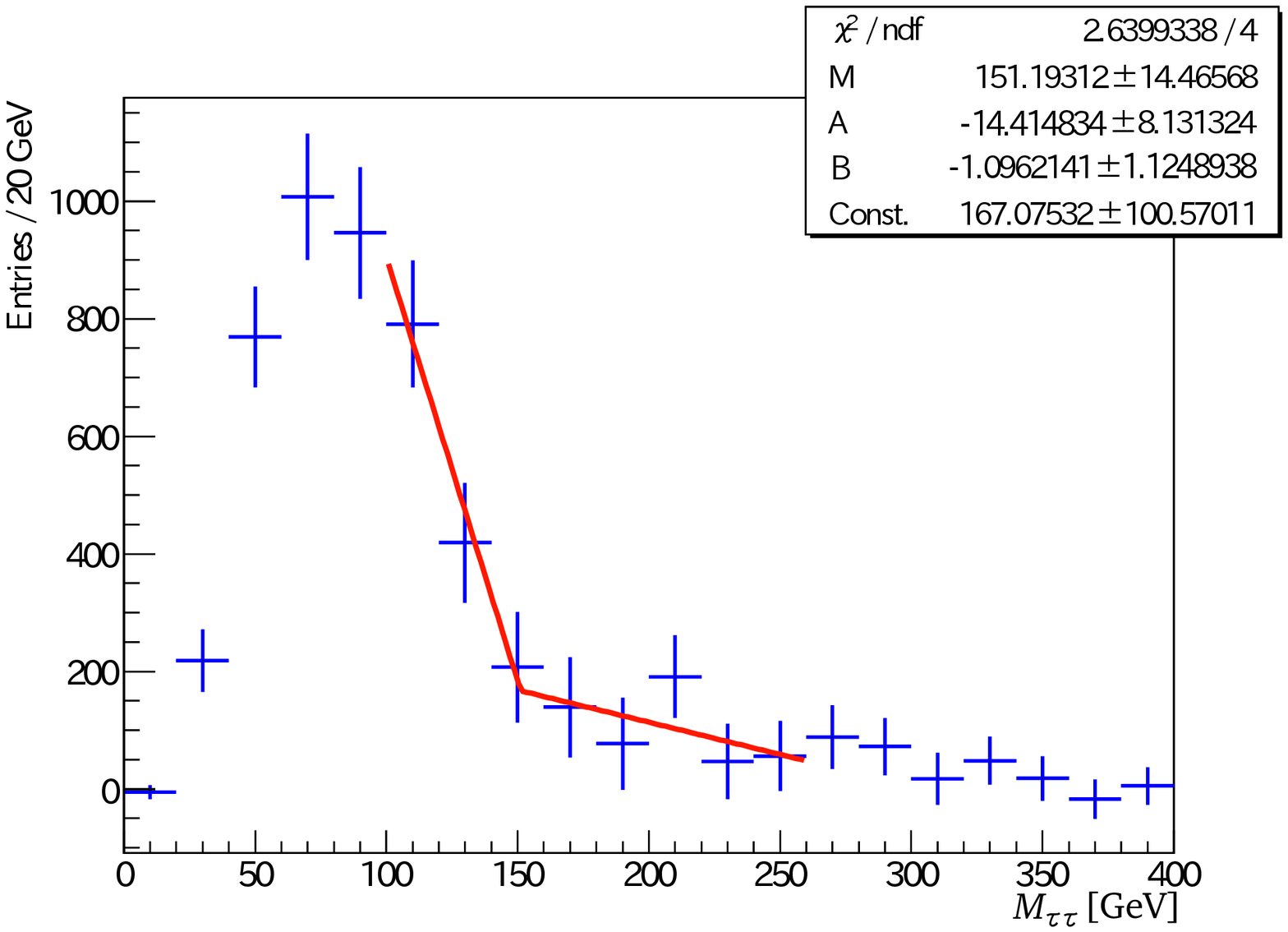} \\
\vspace{0.5cm}
\includegraphics[origin=b, angle=0, width=8.4cm]{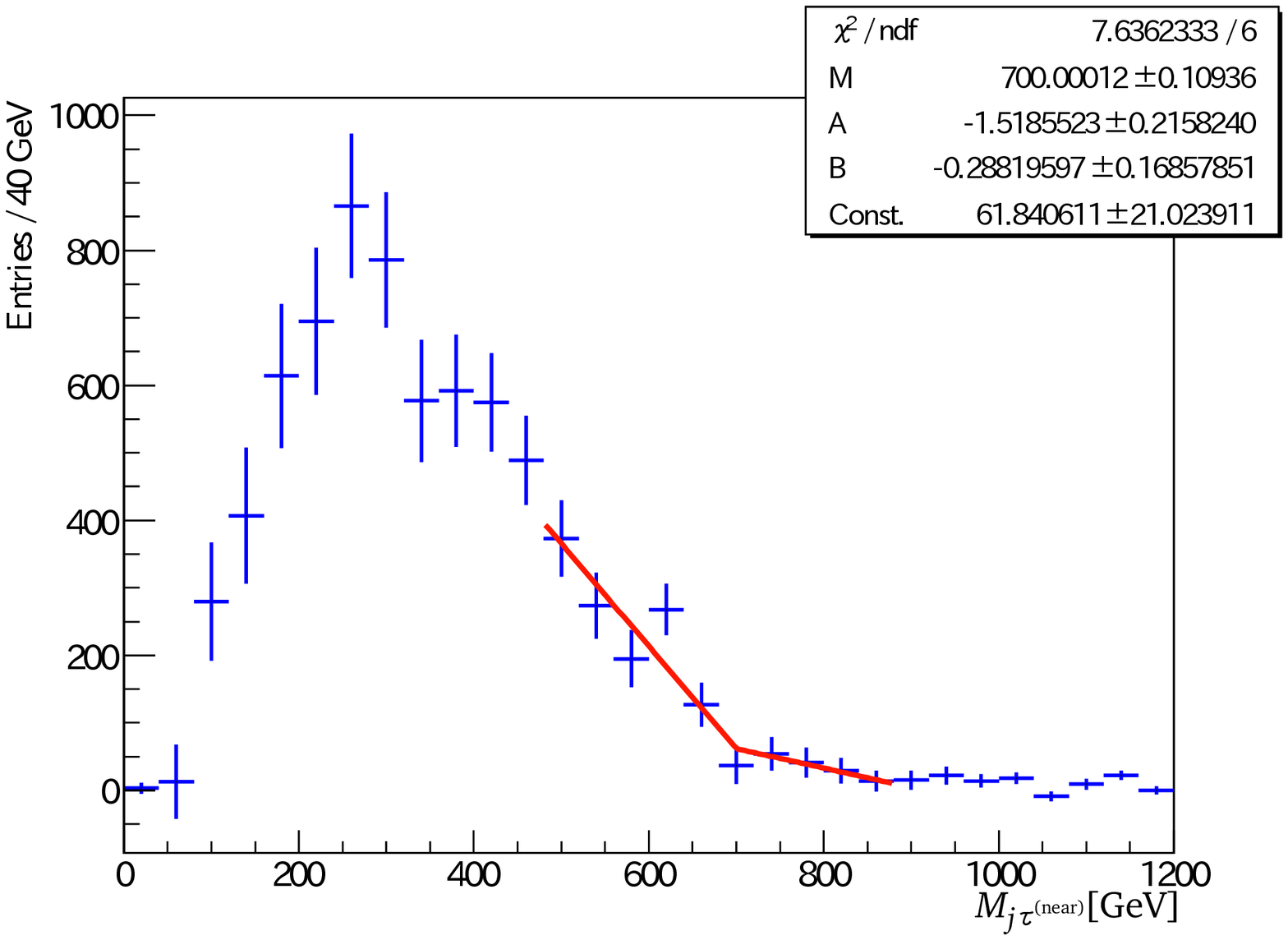} \\
\vspace{0.5cm}
\includegraphics[origin=b, angle=0, width=8.4cm]{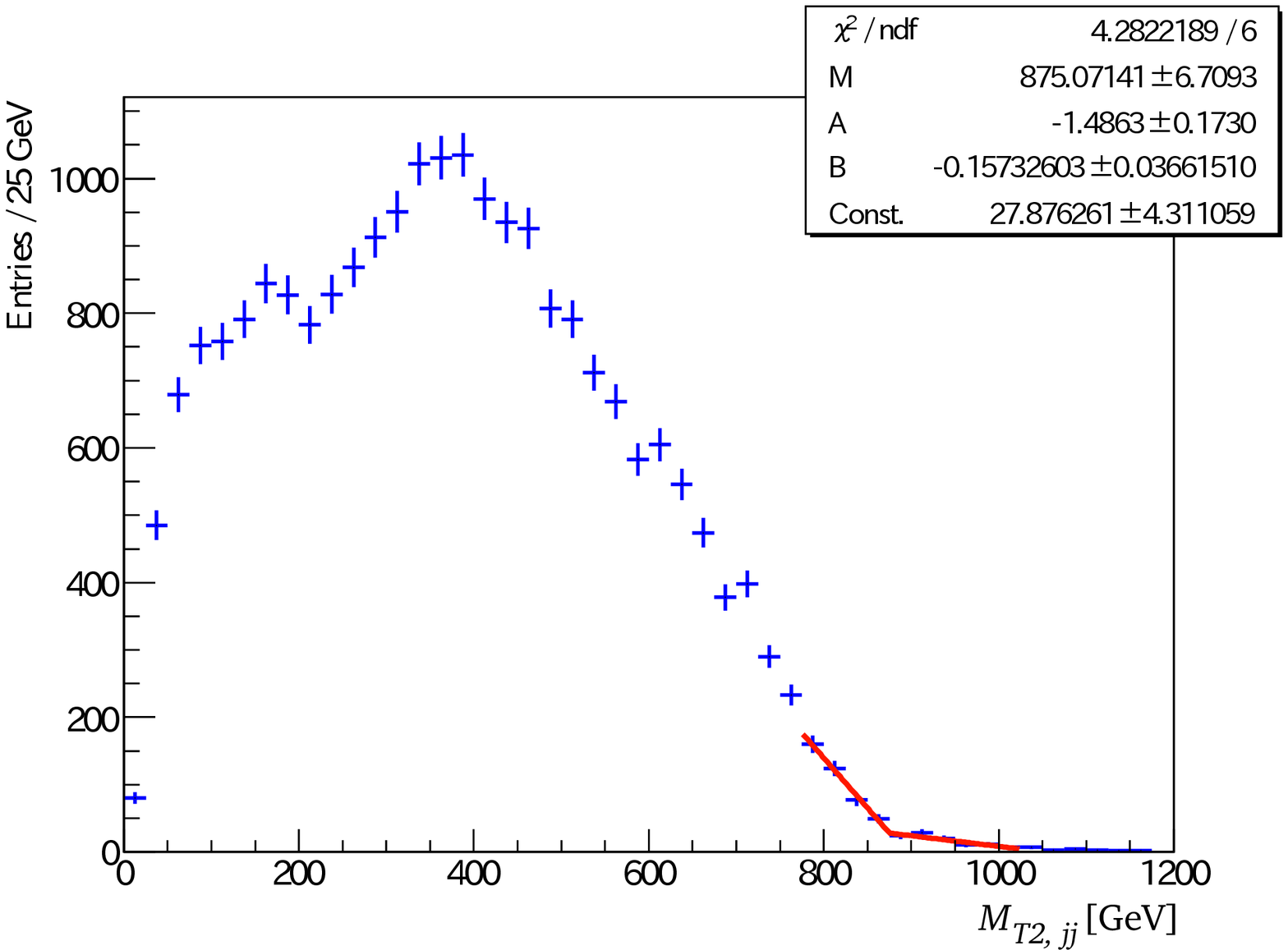}
\caption{\small (Upper panel) Distribution of the invariant mass
  between two tau-jets, $M_{\tau\tau}$. (Middle panel) Distribution of
  the invariant mass between hard jet and near $\tau$-jet,
  $M_{j\tau^{\rm (near)}}$. (Lower panel) Distribution of the
  $M_{T2,jj}$ variable defined in Eq.~(\ref{eq: MT2_jj}) with $m_{\rm
    miss}$ being zero.}
\label{fig: endpoints}
\end{center}
\end{figure}

The first kinematical variable used in the analysis for the mass
spectrum is the invariant mass of two $\tau$-leptons in the decay
chain of a squark. In Fig.~\ref{fig: endpoints} (upper panel), the
distribution of the invariant mass ($M_{\tau\tau}$) after applying the
charge subtraction method is shown. We can see a clear edge at
$M_{\tau\tau} \simeq$ 150 GeV. In order to extract the location of the
endpoint, we use the following fitting function,
\begin{equation}
f(M_{\tau\tau})
=
\left\{
\begin{array}{ll}
A (M_{\tau\tau} - M_{\tau\tau}^{\rm fit}) + C & : M_{\tau\tau} < M_{\tau\tau}^{\rm fit} \\
B (M_{\tau\tau} - M_{\tau\tau}^{\rm fit}) + C & : M_{\tau\tau} > M_{\tau\tau}^{\rm fit}
\end{array}
\right.,
\label{eq: fitting function}
\end{equation}
where $A$, $B$, $M_{\tau\tau}^{\rm fit}$, and $C$ are parameters to
fit the shape of the distribution around the endpoint. With the use of
this bilinear function for the fitting, the location of the endpoint
is determined to be $M_{\tau \tau}^{\rm fit} = 151.2 \pm
14.5$GeV. Notice that the underlying value (the input value on the
simulation) is 151.6 GeV.

\subsubsection{Endpoint on $M_{q \tau^{\rm (near)}}$}

Second kinematical variable we use is the invariant mass between
$\tau$-lepton and jet emitted by the decay of a squark. Using
information about the impact parameter, it is possible to distinguish
near and far $\tau$-leptons with high efficiency. For each
($\tau$,~$\tau$)-pair, we identify the $\tau$-jet whose track has a
larger impact parameter than the other as the far tau-jet, while the
$\tau$-jet with a smaller impact parameter is regarded as the near
$\tau$-jet. After this identification, both combinations of $(j_1,
\tau^{\rm (near)})$ and $(j_2,~\tau^{\rm (near)})$ are used to
calculate $M_{q \tau^{({\rm near})}}$. In the analysis, we also
require that the pair of two tau-jets should satisfy $M_{\tau\tau} <
M_{\tau \tau}^{\rm fit} =$ 155.72GeV in order to reduce fake-QCD and
combinatorial backgrounds.

The distribution of the invariant mass ($M_{q \tau^{\rm (near)}}$)
after applying the charge subtraction method is shown in
Fig.~\ref{fig: endpoints} (middle panel). The endpoint is, again,
fitted by using the bilinear function given in Eq.~(\ref{eq: fitting
  function}). It then turns out that the location of the endpoint on
$M_{q \tau^{\rm (near)}}$ is $M_{q \tau^{\rm (near)}}^{\rm fit} =
700.0 \pm 0.1$GeV. Notice that the underlying value is now 692.3GeV.

\subsubsection{Endpoint on $M_{T2,jj}$}

The last kinematical variable is $M_{T2,jj}$ defined by two hard jets
as visible particles, as mentioned in section \ref{sec: IP for
  masses}. In our analysis, we take the test mass ($m_{\rm miss}$) in
Eq.~(\ref{eq: MT2_jj}) to be zero, so that the endpoint of this
kinematical variable gives $M_{T2,jj}^{\rm max}(0) = (m_{\tilde{q}}^2
- m_{\tilde{\chi}_1^0}^2) / m_{\tilde{q}}$. The distribution of
$M_{T2,jj}(0)$ is shown in Fig.~\ref{fig: endpoints} (lower panel). It
can be seen that a very clear endpoint exists at $M_{T2,jj}(0) \simeq
870$GeV. As in the cases of previous kinematical variables, we fit the
shape of the distribution around the endpoint by the bilinear
function. The endpoint of the distribution is then obtained as
$M_{T2,jj}^{\rm fit}(0) = 875.1 \pm 6.7$GeV (the underlying value is
879.8GeV).

\subsubsection{Mass determination}

When the gravitino mass is neglected, the masses of sparticles,
$m_{\tilde{q}}$, $m_{\tilde{\chi}_1^0}$, and $m_{\tilde{\tau}_1}$, are
determined by three kinematical endpoints of the variables, $M_{\tau
  \tau}^{\rm max}$, $M_{q \tau^{\rm (near)}}^{\rm max}$, and
$M_{T2,jj}^{\rm max}(0)$. Using analytic expressions for the these
endpoints shown in eqs.(\ref{eq: mass_tautau}), (\ref{eq: mass_jtau}),
and (\ref{eq: MT2_jj}), the masses of sparticles are determined by
minimizing the following $\chi^2$ function,
\begin{equation}
\chi_{\rm m}^2
=
\left[
\frac{M_{\tau \tau}^{\rm max} - M_{\tau \tau}^{\rm fit}}
{\Delta M_{\tau \tau}^{\rm fit}}
\right]^2
+
\left[
\frac{M_{q \tau^{\rm (near)}}^{\rm max} - M_{q \tau^{\rm (near)}}^{\rm fit}}
{\Delta M_{q \tau^{\rm (near)}}^{\rm fit}}
\right]^2
+
\left[
\frac{M_{T2,jj}^{\rm max}(0) - M_{T2,jj}^{\rm fit}(0)}
{\Delta M_{T2,jj}^{{\rm fit}}(0)}
\right]^2,
\label{eq:chi2_mass}
\end{equation}
where $M^{\rm fit}$ denotes the center value of the measured endpoint,
and $\Delta M^{\rm fit}$ is its (statistical) error. After minimizing
$\chi^2_{\rm m}$ by varying the input values, $m_{\tilde{q}}$,
$m_{\tilde{\chi}_1^0}$, and $m_{\tilde{\tau}_1}$, we obtain following
results; the right-handed squark mass is $m_{\tilde{q}} = 915.9 \pm
6.4$GeV (the true value is 922.0GeV), the lightest neutralino mass is
$m_{\tilde{\chi}^0_1} = 193.4 \pm 19.5$GeV (the true value is
197.3GeV), and the lightest stau mass is $m_{\tilde{\tau}_1} = 120.5
\pm 18.1$GeV (the true value is 126.2GeV).  Here, the gravitino mass
is taken to be zero (i.e., negligibly small). 

\subsection{Lifetime of the stau NLSP}   \label{sec: results lifetime}

The strategy to determine the lifetime of the stau NLSP is the use of
the distribution of the transverse impact parameter. After showing the
distribution for several input values of $c\tau_{\tilde \tau}$, we
discuss how accurately the lifetime can be determined at the LHC.

\subsubsection{Distribution of the impact parameter}

\begin{figure}[t]
\begin{center}
\includegraphics[origin=b, angle=0, width=8.4cm]{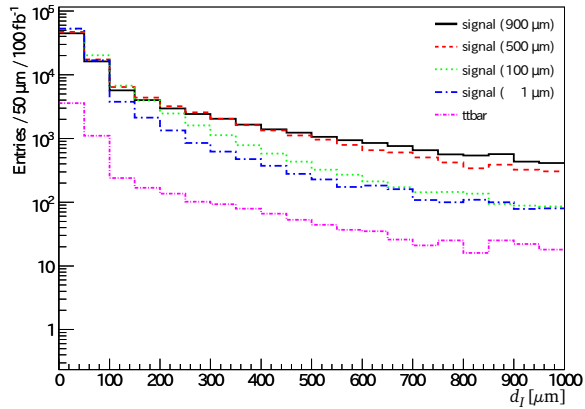}
\caption{\small Distribution of the transverse impact parameter of
  $\tau$-jets after applying the basic kinematical cuts in previous
  section. We show distributions in the SUSY model with $c\tau_{\tilde
    \tau} =$ 1, 100, 500 and 900$\mu$m, respectively, and also show
  the distribution in $t\bar{t}$ production.}
\label{fig: di}
\end{center}
\end{figure}

Distribution of the transverse impact parameter ($d_I$) obtained from
hadronically decays of $\tau$-leptons is shown in Fig.~\ref{fig: di}
with the use of generated events which are passed through the
kinematical cuts discussed in previous section. Four distributions are
shown in this figure with choices of the decay length of the stau NLSP
to be $c\tau_{\tilde \tau} =$ 1, 100, 500, and 900$\mu$m,
respectively. It is clearly seen that the distribution depends on the
decay length of the stau NLSP as expected.

It is also seen that a number of events are found in inner bins with
$d_I \ll c\tau_{\tilde \tau}$ and that a broad tail-structure exists in
the region $d_I \gg c\tau_{\tilde \tau}$. In fact, in both regions,
backgrounds are expected to contribute to the distribution. In small
$d_I$ region, the distribution is dominated by background $\tau$-jets
such as QCD-originated fake ones. Though most of those backgrounds do
not have finite $d_I$ at the parton-level, the backgrounds acquire
finite values of $d_I$ at the detector-level because of the limited
resolution for the vertexing. On the other hand, in the region of $d_I
\gg c\tau_{\tilde \tau}$, backgrounds come from decays of heavy
hadrons such as $D$ or $B$ mesons.
Those hadrons sometimes produce fake $\tau$-jets after flying a sizable distance.
In order to eliminated these backgrounds efficiently, we vary and optimize the
upper and lower endpoints of the bins which are used for the $\chi^2$
analysis as we change $c\tau_{\tilde \tau}^{\rm (test)}$ (where
$c\tau_{\tilde \tau}^{\rm (test)}$ is the test value of the decay
length used to generate a template of $d_I$ distribution).

\subsubsection{Lifetime estimation}

According to the strategy to estimate the lifetime discussed in
section \ref{sec: IP for lifetime}, we now study how well we can
constrain the lifetime of $\tilde{\tau}_1$.  First, we use the
template generated with the true value of the averaged velocity of the
stau NLSP, which is $\bar{\beta}_{\tilde{\tau}} = 0.88 c$ in our
representative point.  In order to see how the result depends on the
underlying value of the lifetime of $\tilde{\tau}_1$, here we use
several values of $c\tau_{\tilde \tau}$ in generating the
impact-parameter distribution.  In addition, in preparing the
templates for $d_I$ distributions, TAUOLA library~\cite{TAUOLA} is used
to simulate the $\tau$ decay event, which enable us to deal with
chirality and finite lifetime of $\tau$-leptons. The range of the test
value is taken to be 10$\mu$m--1100$\mu$m every 10$\mu$m.

With the use of the templates, we perform $\chi^2$-analysis to
determine the lifetime.  In our analysis, only $\tau$-jets satisfying
$0.5 \times c\tau_{\tilde \tau}^{\rm (test)} < d_I < 2.0 \times
c\tau_{\tilde \tau}^{\rm (test)}$ with small pseudo-rapidity, $|\eta|
< 1.0$, are adopted. The size of bin used in each distribution is set
to be $0.15 \times c\tau_{\tilde \tau}^{\rm (test)}$ and, as a result,
we have 10 bins in total. The $\chi^2$ variable to estimate the
lifetime of the stau NLSP is therefore given by
\begin{equation}
\chi^2({c\tau_{\tilde \tau}^{\rm (test)}})
\equiv
\sum_{i = 1}^{10}
\left[
\frac{N^{\rm (th)}_i(c\tau_{\tilde \tau}^{\rm (test)})
-
N^{\rm (exp)}_i}{\Delta N_i}
\right]^2,
\end{equation}
where $N^{\rm (th)}_i(c\tau_{\tilde \tau}^{\rm (test)})$ denotes the
number of signals in the $i$-th bin obtained by the template for a
given $ c\tau_{\tilde \tau}^{\rm (test)}$, while $N^{\rm (exp)}_i$ is
the one obtained by using generated events. We only involves the
statistical error as $\Delta N_i \equiv \sqrt{N^{\rm (exp)}_i}$ in the
analysis. The degrees of the freedom in this $\chi^2$-test is
therefore $(10 - 1) = 9$, and the hypothesis is excluded at 95\%
C.L. when $\chi^2 > 16.92$.

\begin{figure}[t]
\begin{center}
\includegraphics[origin=b, angle=0,width=8.4cm]{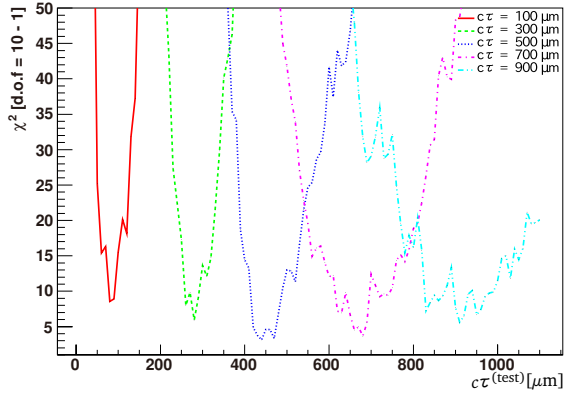}
\caption{\small $\chi^2$-values as a function of the test lifetime
  (the test decay length) of the stau NLSP, $c\tau_{\tilde \tau}^{\rm
    (test)}$, for the underlying values of $c\tau_{\tilde \tau} =$ 100
  (red, solid), 300 (green, dashed), 500 (blue, dotted), 700 (violet,
  dot-dashed), and 900 (cyan, dot-dot-dashed) $\mu$m.
  Here, the averaged velocity is taken to be 
  $\bar{\beta}_{\tilde{\tau}}=0.88$.}
\label{fig: chi2}
\end{center}
\end{figure}

Resultant $\chi^2$-values as a function of the test lifetime (the test
decay length) of the stau NLSP, $c\tau_{\tilde \tau}^{\rm (test)}$,
for underlying values of $c\tau_{\tilde \tau} =$ 100 (red, solid), 300
(green, dashed), 500 (blue, dotted), 700 (violet, dot-dashed), and 900
(cyan, dot-dot-dashed) $\mu$m are shown in Fig.~\ref{fig: chi2}. From
these results, the lifetime (the decay length) of the stau NLSP in
each case is determined at 95\% C.L. to be
\begin{eqnarray}
\begin{array}{rcll}
50\mu{\rm m} & \lesssim~c\tau_{\tilde \tau}~\lesssim & 110\mu{\rm m}
& {\rm (underlying}~c\tau_{\tilde \tau} = 100\mu{\rm m}), \\
240\mu{\rm m} & \lesssim~c\tau_{\tilde \tau}~\lesssim & 330\mu{\rm m}
& {\rm (underlying}~c\tau_{\tilde \tau} = 300\mu{\rm m}), \\
410\mu{\rm m} & \lesssim~c\tau_{\tilde \tau}~\lesssim & 540\mu{\rm m}
& {\rm (underlying}~c\tau_{\tilde \tau} = 500\mu{\rm m}), \\
570\mu{\rm m} & \lesssim~c\tau_{\tilde \tau}~\lesssim & 800\mu{\rm m}
& {\rm (underlying}~c\tau_{\tilde \tau} = 700\mu{\rm m}), \\
810\mu{\rm m} & \lesssim~c\tau_{\tilde \tau}~\lesssim & 1060\mu{\rm m}
& {\rm (underlying}~c\tau_{\tilde \tau} = 900\mu{\rm m}). \\
\end{array}
\end{eqnarray}
We can see that, if the correct value of $\bar{\beta}_{\tilde{\tau}}$
is used, the analysis based on the impact parameter distribution gives
a good estimate of the lifetime (the decay length) of the stau NLSP
with accuracy of about 30\% when $c\tau_{\tilde \tau} > 100\mu$m. If
more precise information about the velocity distribution of
$\tilde{\tau}_1$ is available, better estimate of the lifetime may be
obtained.  

So far, we have neglected the uncertainty arising from the
determination of the velocity distribution of $\tilde{\tau}_1$.  As we
have mentioned, the detailed study of the uncertainty in the velocity
distribution is beyond the scope of this paper. In our analysis,
however, we estimated the uncertainty of the
$\bar{\beta}_{\tilde{\tau}}$ determination related to the errors in
mass measurements and also to the production process in order to
demonstrate that the $\bar{\beta}_{\tilde{\tau}}$ can be obtained with
some accuracy.

First, in order to study the effects of the errors in the mass
measurements, we generated the events using the different sparticle
mass spectrum from our representative point. Here, we used the mass
spectrum predicted from the simple gauge mediation model except for
$\tilde{\tau}_1$ because we found that the error of
$m_{\tilde{\tau}_1}$ is the largest among the reconstructed masses in
the previous subsection. Then we generated the full SUSY events by
varying $m_{\tilde{\tau}_1}$ by $\pm 20$GeV, and found that
the value of $\bar{\beta}_{\tilde{\tau}}$ changes by $\sim 0.01$.
In addition, the dominant SUSY process may not be well understood in
the actual situation.  If so, it may be reasonable to estimate
$\bar{\beta}_{\tilde{\tau}}$ by assuming the process we use for our
analysis, which is the process shown in Fig.\ \ref{fig: cascade
  chain}.  We generated events corresponding to such a process (using
the correct mass relation).  Then, $\bar{\beta}_{\tilde{\tau}}$ is
found to be $\sim 0.93$.  Thus, a relatively larger uncertainty of
$\Delta\bar{\beta}_{\tilde{\tau}}\sim 0.05$ is expected if the
dominant SUSY process cannot be understood.

\begin{figure}[t]
\begin{center}
\includegraphics[origin=b, angle=0,width=8.4cm]{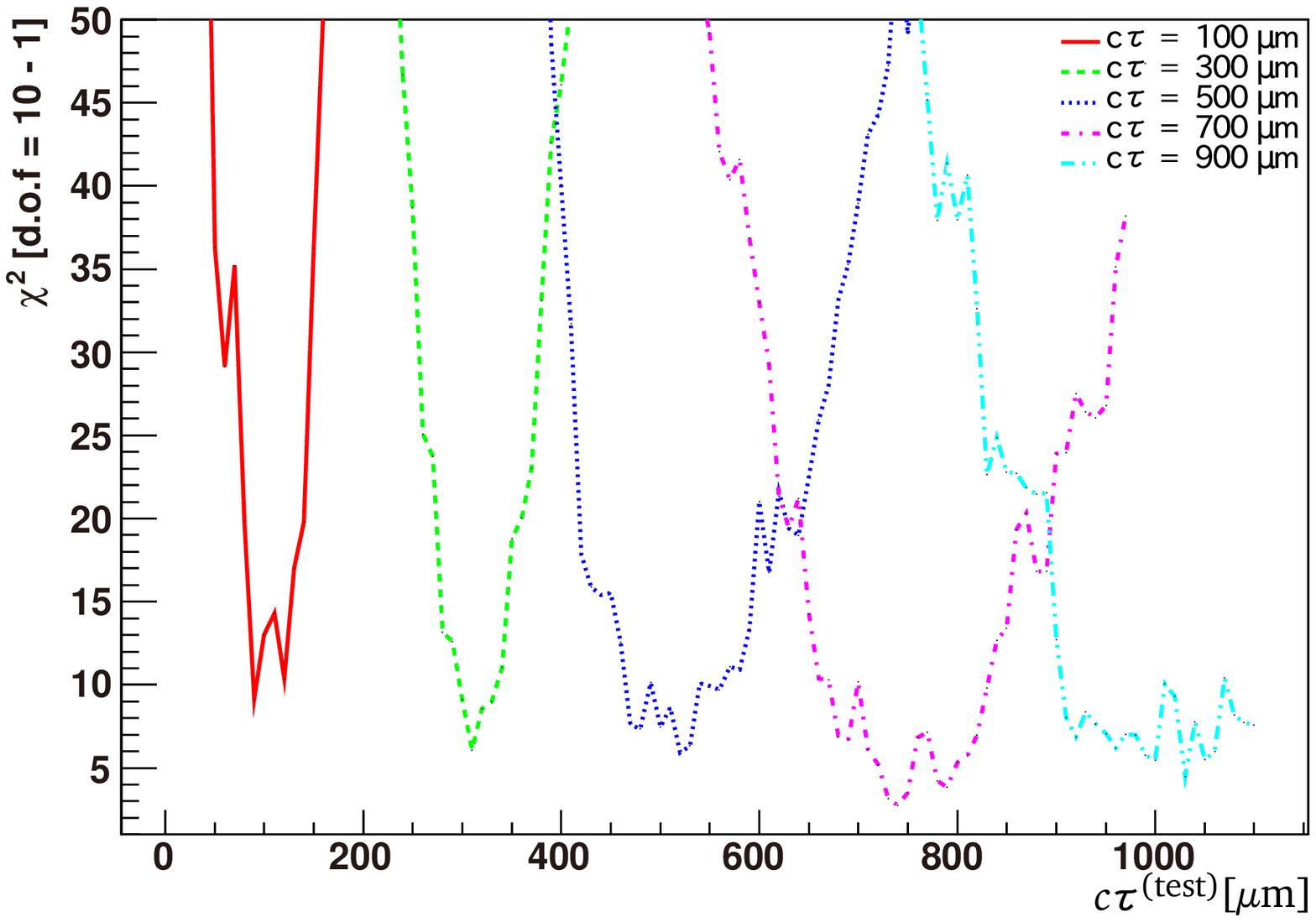}
\caption{\small Same as Fig.~\ref{fig: chi2}, except for
  $\bar{\beta}_{\tilde{\tau}}=0.83$.}
\label{fig:chi2_0.83}
\end{center}
\begin{center}
\includegraphics[origin=b, angle=0,width=8.4cm]{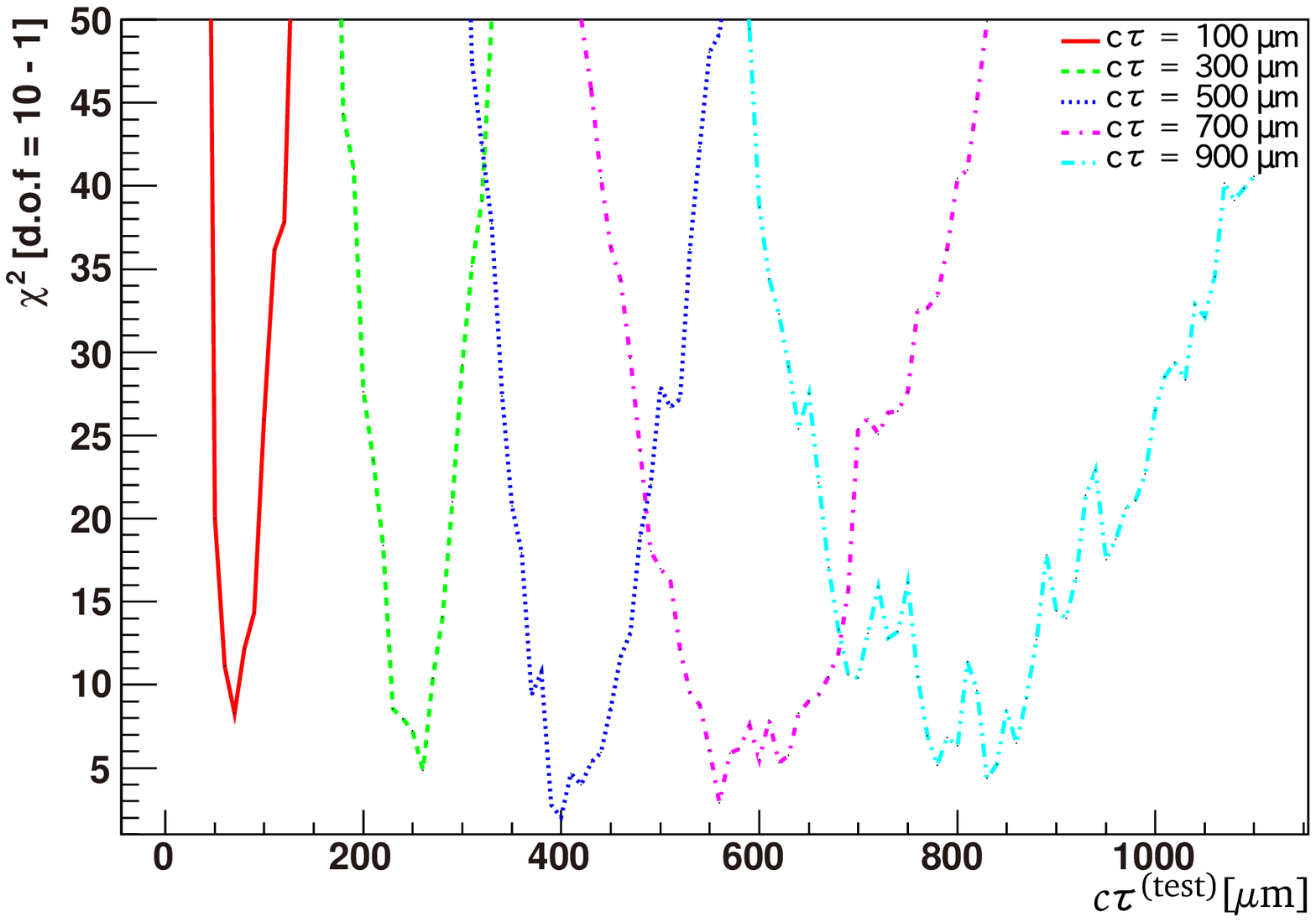}
\caption{\small Same as Fig.~\ref{fig: chi2}, except for
  $\bar{\beta}_{\tilde{\tau}}=0.93$.}
\label{fig:chi2_0.93}
\end{center}
\end{figure}

To see how this affects the determination of the lifetime, we
calculate the $\chi^2$ variable using the template with
$\bar{\beta}_{\tilde{\tau}}=0.83$ and $0.93$.  The results are shown
in Figs.~\ref{fig:chi2_0.83} and \ref{fig:chi2_0.93}.  We can see
that error related to the uncertainty to the averaged velocity is
$\sim 50-100\mu{\rm m}$ if $\Delta\bar{\beta}_{\tilde{\tau}}\sim
0.05$.  Even with such an uncertainty, we can still have a relatively
good determination of $\tau_{\tilde \tau}$.  Thus, the impact
parameter will be a powerful tool to measure the lifetime at the LHC.

Finally, we consider how well we can estimate the gravitino mass if we
assume that $\tilde{\tau}_1$ decays into the gravitino and $\tau$-lepton.
In the case with the underlying gravitino mass of 9.7eV,
we have shown that the lifetime of the stau is estimated between 410$\mu$m and 540$\mu$m for $\bar{\beta}_{\tilde{\tau}}=$~0.88.
In addition, the uncertainty related to $\bar{\beta}_{\tilde{\tau}}$ is estimated to be about 40$\mu$m, as one can see from Figs.~\ref{fig:chi2_0.83} and~\ref{fig:chi2_0.93}.
Then, by using the measured stau mass $m_{\tilde{\tau}} = 120.5 \pm 18.1$GeV and the center value of the estimated NLSP lifetime $c\tau_{\tilde{\tau}} = 475\mu$m,
we obtain $m_{3/2} = 8.3\pm 1.8\pm 0.8\pm 1.0$eV, where
the errors originate in the uncertainties of $m_{\tilde{\tau}}$, $c\tau_{\tilde{\tau}}$ and $\bar{\beta}_{\tilde{\tau}}$, respectively.

\section{Summary}
\label{sec: summary}

We have proposed a method to determine the mass spectrum of sparticles
and the lifetime of the stau NLSP at the LHC when the mass of the
gravitino LSP is of the order of 10eV. Though the decay length of the
stau NLSP is very short, which is of the order of 100--1000$\mu$m, it
is still possible to deeply study the model by utilizing the
transverse impact parameter of tau-jets from the decay of the stau
NLSP.

We have first discussed the mass measurement of sparticles using a
typical cascade decay chain of a squark shown in Fig.~\ref{fig:
  cascade chain}. This SUSY event involves, at least, four
$\tau$-leptons, which makes it difficult to analyze the signal
because of combinatorial backgrounds. Information about the impact
parameter of the tau-jet, however, resolves the problem, and we have
shown that the mass spectrum of sparticles can be determined
accurately through the kinematical endpoints of $M_{\tau \tau}$, $M_{j
  \tau^{ \rm (near)} }$, and $M_{T2,jj}$.

We have also discussed the determination of the lifetime of the stau
NLSP using the distribution of the transverse impact parameter. The
impact parameter depends not only on the lifetime but also on the
velocity of the stau NLSP. We have therefore developed a strategy to
estimate the velocity by utilizing a simulation with information about
the mass spectrum obtained in the previous stage.  We have shown that,
if the velocity distribution of $\tilde{\tau}_1$ is somehow
understood, the lifetime of the stau NLSP is determined with the
accuracy of about 30\% as far as its decay length is larger than
$\sim 100 \mu$m.
Thus, if the underlying model of the SUSY breaking is low-scale gauge mediation with the gravitino mass of ${\cal O}(10)$eV, the LHC may have a chance to acquire some information
about the gravitino mass.

\section*{Acknowledgments}

This work is supported by Grant-in-Aid for Scientific research from
the Ministry of Education, Science, Sports, and Culture (MEXT), Japan,
Nos.\ 21740174 \& 22244031 (S.M.), No.\ 22540263 (T.M.), and No.\
22244021 (M.A., S.M. and T.M.), by World Premier International
Research Center Initiative (WPI Initiative), MEXT, Japan,
and by JSPS Research Fellowships for Young Scientists, MEXT, Japan (T.I.).
M.A. also acknowledges support from the German Research Foundation (DFG) through
grant BR 3954/1-1


\begin{thebibliography}{99} 


\bibitem{BookDrees}
See, for example,
M.~Drees, R.~M.~Godbole and P.~Roy,
{\it Theory and phenomenology of sparticles},
(World Scientific, 2004).


\bibitem{Dine:1993yw}
  M.~Dine, A.~E.~Nelson,
  Phys.\ Rev.\  {\bf D48}, 1277 (1993),
%
  M.~Dine, A.~E.~Nelson, Y.~Shirman,
  Phys.\ Rev.\  D {\bf 51}, 1362 (1995),
%
  M.~Dine, A.~E.~Nelson, Y.~Nir, Y.~Shirman,
  Phys.\ Rev.\  D {\bf 53}, 2658 (1996).

\bibitem{Feng:2010ij}
J.~L.~Feng, M.~Kamionkowski and S.~K.~Lee,
Phys.\ Rev.\  D {\bf 82}, 015012 (2010).

\bibitem{Kawasaki:2008qe}
  M.~Kawasaki, K.~Kohri, T.~Moroi and A.~Yotsuyanagi,
  Phys.\ Rev.\  D {\bf 78} (2008) 065011.

\bibitem{Viel:2005qj}
  M.~Viel, J.~Lesgourgues, M.~G.~Haehnelt, S.~Matarrese, A.~Riotto,
  Phys.\ Rev.\  {\bf D71}, 063534 (2005).

\bibitem{Matsumoto:2011fk}
  S.~Matsumoto and T.~Moroi,
  Phys.\ Lett.\  B {\bf 701} (2011) 422.


\bibitem{Aad:2009wy}
  G.~Aad {\it et al.}  [The ATLAS Collaboration],
  arXiv:0901.0512 [hep-ex].

\bibitem{Bayatian:2006zz}
  G.~L.~Bayatian {\it et al.}  [CMS Collaboration],
  CMS-TDR-008-1 (2006).

\bibitem{Ishiwata:2008tp}
  K.~Ishiwata, T.~Ito and T.~Moroi,
  Phys.\ Lett.\  B {\bf 669}, 28 (2008).
\bibitem{Kaneko:2008re}
  S.~Kaneko, J.~Sato, T.~Shimomura, O.~Vives and M.~Yamanaka,
  Phys.\ Rev.\  D {\bf 78}, 116013 (2008).
\bibitem{Asai:2008sk}
  S.~Asai, T.~Moroi and T.~T.~Yanagida,
  Phys.\ Lett.\  B {\bf 664}, 185 (2008)
  [arXiv:0802.3725 [hep-ph]].
\bibitem{Asai:2011wy}
  S.~Asai, Y.~Azuma, M.~Endo, K.~Hamaguchi and S.~Iwamoto,
  arXiv:1103.1881 [hep-ph].

\bibitem{Kats:2011qh}
  See, Y.~Kats, P.~Meade, M.~Reece, D.~Shih, 
  [arXiv:1110.6444 [hep-ph]] and references therein.

\bibitem{Hinchliffe:1996iu}
  See, for example,
  I.~Hinchliffe, F.~E.~Paige, M.~D.~Shapiro, J.~Soderqvist, W.~Yao,
  Phys.\ Rev.\  {\bf D55}, 5520-5540 (1997).

\bibitem{Hinchliffe:1998ys}
  I.~Hinchliffe, F.~E.~Paige,
  Phys.\ Rev.\  {\bf D60}, 095002 (1999).


\bibitem{Lester:1999tx}
  C.~G.~Lester and D.~J.~Summers,
  Phys.\ Lett.\  B {\bf 463} (1999) 99.


\bibitem{ISAJET}
  F.~E.~Paige, S.~D.~Protopopescu, H.~Baer and X.~Tata,
  arXiv:hep-ph/0312045.

\bibitem{HERWIG}
  G.~Corcella {\it et al.},
  JHEP {\bf 0101}, 010 (2001);
%
  G.~Corcella {\it et al.},
  arXiv:hep-ph/0210213.

\bibitem{HERWIGSUSY}
  S.~Moretti, K.~Odagiri, P.~Richardson, M.~H.~Seymour and B.~R.~Webber,
  JHEP {\bf 0204}, 028 (2002).

\bibitem{PGS4}
  For information on Pretty Good Simulation of high energy
  collisions (PGS4), see
  {\verb$http://www.physics.ucdavis.edu/%7Econway/research/research.html$}.



\bibitem{TAUOLA}
  S.~Jadach, J.~H.~Kuhn, Z.~Was,
  Comput.\ Phys.\ Commun.\  {\bf 64}, 275-299 (1990),
  N.~Davidson, G.~Nanava, T.~Przedzinski, E.~Richter-Was, Z.~Was,
  [arXiv:1002.0543 [hep-ph]].


\end{thebibliography}
\end{document}